\documentclass[pra]{revtex4-1}

\usepackage{hyperref}
\usepackage[version=3]{mhchem}

\begin{document}

\title{Generalization of the Kohn-Sham system that can represent arbitrary
       one-electron density matrices}
\author{Hubertus J. J. van Dam}
\affiliation{Brookhaven National Laboratory, Upton, NY 11973-5000}
\date{April 26, 2016}

\begin{abstract}
Density functional theory is currently the most widely applied method in 
electronic structure theory. The Kohn-Sham method, based on a fictitious
system of non-interacting particles, is the workhorse of the theory. The 
particular form of the Kohn-Sham wave function admits only idem-potent 
one-electron density matrices whereas wave functions of correlated electrons in 
post-Hartree-Fock methods invariably have fractional occupation numbers.
Here we show that by generalizing the orbital concept, and introducing a
suitable dot-product as well as a probability density a non-interacting
system can be chosen that can represent the one-electron density matrix
of any system, even one with fractional occupation numbers. This fictitious
system ensures
that the exact electron density is accessible within density functional theory.
It can also serve as the basis for reduced density matrix functional 
theory. Moreover, to aid the analysis of the results the orbitals may be 
assigned energies from a mean-field Hamiltonian. This produces energy levels
that are akin to Hartree-Fock orbital energies such that conventional 
analyses based on Koopmans theorem are available. Finally,
this system is convenient in formalisms that depend on 
creation and annihilation operators as they are trivially applied to 
single-determinant wave functions.
\end{abstract}

\maketitle

\section{Introduction}

The dominant approach in electronic structure theory is density functional 
theory (DFT). Its foundations were laid with the Hohenberg-Kohn 
theorems~\cite{Hohenberg1964}. The theory was turned into a practical approach
by the Kohn-Sham method~\cite{Kohn1965}. A central component of the method
is a fictitious system of non-interacting electrons. The wave function that
represents this system is supposed to be able to generate the exact electron
density. In practice the Kohn-Sham wave function has the same form as the
Hartree-Fock wave function. This wave function can only generate idem-potent 
density matrices whereas all post-Hartree-Fock methods, including Full-CI,
generate density matrices with fractional occupation numbers.
In this paper we show that is possible to generalize the Kohn-Sham wave function
such that it can generate any one-electron density matrix. At the same
time this generalized wave function remains representative of a system of
non-interacting electrons, i.e. a single-determinant wave function.
Incidentally, this wave function can also serve as a basis for reduced density
matrix functional theory (RDMFT). This theory is an extension of DFT to
density matrices formulated by Gilbert~\cite{Gilbert1975}. A particular 
feature of Gilbert's theory is that at convergence all orbitals with fractional
occupation numbers are degenerate. In this work that requirement results
in all correlated electrons being degenerate.

An interesting aspect of this wave function is that as it is a 
single-determinant wave function it is also possible to formulate one-electron 
properties. For instance it is possible to calculate single-electron energy
levels akin to Hartree-Fock energies even if the wave function is a solution
of RDMFT. Furthermore, it is trivial to apply annihilation and creation 
operators to a single-determinant wave function. Hence it is easy to generate 
excited determinants as well.

In the remainder of the paper Section~\ref{sect:def-wfn} defines the generalized
wave function. It is shown that the single-determinant wave function is indeed
capable of generating an arbitrary one-electron density matrix. 
Section~\ref{sect:opt-wfn} explains how the wave function may be optimized.
It is shown that the equations are actually in part the same as the Kohn-Sham
equations, only another set of very similar equations is needed to describe
the correlated electrons. Section~\ref{sect:wfn-ana} describes how the
wave function can be canonicalized to yield a Hartree-Fock like one-electron
picture. Finally, section~\ref{sect:app} shows some simple applications to
demonstrate the features of the approach proposed.

\section{Defining the wave function}
\label{sect:def-wfn}

The approach outlined in this article addresses the fact that for an 
$n_b$-dimensional density matrix there does not exist an
$n_b$-dimensional transformation that only changes the occupation 
numbers. Nevertheless, such a transformation can be effected by mapping
the problem into an $n_b^2$-dimensional space, applying suitably chosen
rotations, and projecting the result back into the original $n_b$-dimensional
space. 
In this section the definition of an orbital is generalized such that it maps
the density matrix from an $n_b$-dimensional to an $n_b^2$-dimensional space.
In addition the probability density is choosen to generate the projection from
the $n_b^2$-dimensional density matrix back to an $n_b$-dimensional one.
An illustrative example of these operations is provided in the Supplementary
Material~\cite{vanDamsupplement}. Also a dot-product for the generalized
orbitals is chosen.

It is shown that the generalized orbitals form an orthonormal set of
the same dimension as the conventional orbitals. It is also shown that the 
density matrix of a single generalized orbital matches the appropriate 
N-presentability conditions~\cite{Coleman1963} whereas it may distribute
the electron over any 
set of conventional orbitals. Finally it is shown that these generalized
orbitals
may be used to define a single Slater determinant wave function. Combining
the orthonormality as well as the density matrix implied in the probability
density it is shown that the resulting Slater determinant can generate any 
one-electron density matrix, even one of a correlated state, i.e. a
non-idempotent one.

The arguments in this section are most easily formulated in the form of 
matrix-vector equations. Note that the arguments given here can be formulated
for a single spin-channel (either the $\alpha$- or $\beta$-electron channel)
without loss of generality.
First we recall that the one-electron density matrix $D$ for a system
represented in $n_b$ basis functions is a non-negative Hermitian matrix. Hence
it may be diagonalized to produce an orthonormal set of eigenvectors,
conventionally called natural orbitals, represented
as matrix $N$, where each natural orbital is a column of $N$, and eigenvalues 
$1 \ge d_1 \ge d_2 \ge \ldots d_{n_b} \ge 0$. These eigenvalues are referred
to as occupation numbers. The natural
orbitals may be expressed in an orthogonal basis such as plane waves or in 
a non-orthogonal basis such as Gaussian type orbitals. To accommodate all options
an overlap matrix $S$ is introduced, the elements of which are defined as
\begin{eqnarray}
  S_{ab} &=& \left\langle\left.\chi_a(r)\right|\chi_b(r)\right\rangle \\
         &=& \int \chi^*_a(r)\chi_b(r) \mathrm{d}r
\end{eqnarray}
where $a$, $b$, and later $c$ label basis functions, and $\chi_a(r)$ represents
the basis functions.
In the case of an orthogonal basis
the overlap matrix is simply the unit matrix. As the natural orbitals $N$ are an
orthonormal set the following condition holds
\begin{eqnarray}
  \left\langle\left.N_i\right|N_j\right\rangle
            &=& \sum_{a,b=1}^{n_b}N^*_{ai}S_{ab}N_{bj} \\
            &=& \delta_{ij}
\end{eqnarray}
Note that a matrix with one index, such as $N_i$, refers to a column. In this
paper the indices $i$, $j$ and later $k$ label natural orbitals.

To represent the occupation numbers another set of $n_b$ orthonormal vectors is
introduced that are referred to as correlation functions that are columns of 
matrix $C$. The name refers to the fact that fractional occupation numbers that
may be generated by these functions are directly related to electron correlation
as in post Hartree-Fock methods.
The correlation functions being expressed in terms of the 
natural orbitals are always represented in a orthonormal basis. In a system with
$n_e$ electrons of a given spin there are $n_e$ occupied correlation functions
and all others are unoccupied. The contribution from the $r$-th correlation
function to the $i$-th occupation number is simply $C^*_{ir}C_{ir}$.
For the correlation functions the condition
\begin{eqnarray}
  \left\langle\left.C_r\right|C_s\right\rangle
            &=& \sum_{i,j=1}^{n_b}C^*_{ir}I_{ij}C_{js} \\
            &=& \delta_{rs}
\end{eqnarray}
holds. Indices $r$, $s$ and later $t$ label correlation functions.

With the definitions above a set of $n_b$ generalized orbitals $G$ can be
defined where every vector $G_r$ can be expanded as
\begin{eqnarray}
   \left|G_s(r)\right\rangle
   &=& \sum_{a,i=1}^{n_b} G_{ai,s}\left|\chi_a(r)\right\rangle \\
   &=& \sum_{a,i=1}^{n_b} N_{ai}C_{is}\left|\chi_a(r)\right\rangle
   \label{Eq:genorb}
\end{eqnarray}
I.e. formally every generalized orbital has $n_b^2$ coefficients. As every
correlation function has a one-to-one correspondence to a generalized orbital
the indices $r$, $s$ and $t$ will be used to label the latter also.
To define the dot-product between two vectors from $G$ a metric $S^G$ is 
constructed as a matrix by replicating the overlap matrix $S$ $n_b$-times both
horizontally and vertically as
\begin{equation}
   S^G = \left(
   \begin{array}{cccc}
      S      & S      & \ldots & S      \\
      S      & S      & \ldots & S      \\
      \vdots & \vdots & \ddots & \vdots \\
      S      & S      & \ldots & S 
   \end{array}
   \right)
   \label{Eq:metric}
\end{equation}
or equivalently
\begin{equation}
   S^G_{ai,bj} = S_{ab}
\end{equation}
In terms of this metric the dot-product for vectors from $G$ can be written
as
\begin{eqnarray}
  \left\langle\left.G_r\right|G_s\right\rangle
  &=& \sum_{a,i=1}^{n_b}\sum_{b,j=1}^{n_b}G^*_{ai,r} S^G_{ai,bj} G_{bj,s} \\
  &=& \sum_{i,j=1}^{n_b}C^*_{ir}C_{js}
      \left(\sum_{a,b=1}^{n_b}N^*_{ai}S_{ab}N_{bj}\right) \\
  &=& \sum_{i,j=1}^{n_b}C^*_{ir}C_{js}\delta_{ij} \\
  &=& \delta_{rs}
\end{eqnarray}
Hence it is clear that the vectors $G$ form an orthonormal set. As the 
dimension of the basis set is $n_b$ it is also obvious that there can only 
be $n_b$ linearly independent vectors in this set. Note that if the only
concern is to show that the generalized orbitals constructed from a given set
of natural orbitals and correlation functions form an orthonormal set then the
metric of Eq.~\ref{Eq:metric} can equally well be replaced by a block diagonal 
one. The advantage of Eq.~\ref{Eq:metric} is that it also allows calculating
the overlap between generalized orbitals generated from different sets of
natural orbitals and correlation functions. A block diagonal metric would cause
ambiguities in the latter case.

Turning to the probability density given a vector from $G$ we have
\begin{equation}
   \left|G_s(r)\right\rangle\left\langle G_s(r')\right|
   = \sum_{a,b=1}^{n_b}
     \left|\chi_a(r)\right\rangle D^s_{ab}\left\langle\chi_b(r')\right|
\end{equation}
where $D$ is the density matrix. The elements of the density matrix generated
by the single generalized orbital $s$ are given by
\begin{equation}
   D^s_{ab} = \sum_{i=1}^{n_b}N_{ai}C_{is}C^*_{is}N^*_{bi}
\end{equation}
It is clear from the definition that the matrix $D$ is non-negative, and from
the normalization of the vectors $C$ that the trace is $1$. This is exactly
what is required for the one-electron density matrix of a single electron.

Given the vectors $G$ and assuming a system of $n_e$ electrons in a particular
spin channel we can write a single-determinant wave function in terms of these 
vectors as
\begin{equation}
   \Psi(r_1,\ldots,r_{n_e})
   = \left|G_1(r_1)G_2(r_2)\ldots G_{n_e}(r_{n_e})\right|
   \label{Eq:wavefunction}
\end{equation}
From this wave function the one-electron density matrix can be obtained as
\begin{eqnarray}
  \sum_{a,b=1}^{n_b}\left|\chi_a(r_1)\right\rangle D_{ab}
                    \left\langle\chi_b(r'_1)\right|
  &=& n_e \int \Psi(r_1,r_2,\ldots,r_{n_e})\Psi^*(r'_1,r_2,\ldots,r_{n_e})
      \mathrm{d}r_2 \ldots \mathrm{d}r_{n_e} \\
  D_{ab}
  &=& \sum_{i=1}^{n_b}\sum_{s=1}^{n_e} N_{ai}C_{is}C^*_{is}N^*_{bi}
      \label{Eq:densmat} \\
  &=& \sum_{i=1}^{n_b}N_{ai}d_i N^*_{bi}
      \label{Eq:densmat_occ}
\end{eqnarray}
The occupation numbers $d_i$ are obviously non-negative as they are a sum of 
squares. In addition because the correlation functions are an orthonormal set
expressed in an orthonormal basis the matrix $C$ is unitary. This means that
\begin{equation}
   C^* C = C C^* = I
\end{equation}
from which it follows that if only a subset of all columns of $C$ are included
then the diagonal elements $d_i \le 1$. In addition the trace of the density
matrix term of a single vector of $G$ is $1$ so that if $n_e$ such 
vectors are included in the wave function the trace of the density matrix is
$n_e$. Hence we have shown that a single-determinant wave function of the form
presented here
can generate any one-electron density matrix even non-idempotent ones.
Trivially, an idempotent density matrix can be obtained when the correlation
functions are unit vectors. Thus the original Kohn-Sham system is obtained as 
a special case of the wave function proposed.

\section{Optimizing the wave function}
\label{sect:opt-wfn}

In the previous section a single-determinant wave function that can represent
arbitrary one-electron density matrices was proposed. This wave function is
expressed in terms of correlation functions and conventional natural
orbitals. In order to deploy this wave function an approach to optimize it
is required. Obviously this approach is based on an energy minimization.
To keep the results as general as possible we consider energy expressions of
the form
\begin{equation}
   E = E(D^\alpha,D^\beta)
\end{equation}
I.e. the energy is a functional of the one-electron density matrices of both
spin channels. This type of expression encompasses a wide range of energy 
expressions including Hartree-Fock, DFT, and RDMFT. Furthermore, here only the
optimization w.r.t. one of the spin channels is discussed as the results for the
other spin-channel are the same.
Thus we consider the problem
\begin{eqnarray}
   L &=&
   E(D^\alpha(N^\alpha,C^\alpha));D^\beta)
   + \sum_{i,j=1}^{n_b}d_i\lambda^{N^\alpha}_{ij}
     \left(I_{ij}-\sum_{a,b=1}^{n_b}N^{\alpha*}_{ai}S_{ab}N^\alpha_{bj}\right)
   + \sum_{r,s=1}^{n_b}\lambda^{C^\alpha}_{rs}
     \left(I_{rs}-\sum_{i,j=1}^{n_b}C^{\alpha*}_{ir}I_{ij}C^\alpha_{js}\right)
   \\
   && \min_{N^\alpha,C^\alpha,\lambda^{N^\alpha}_{ij},\lambda^{C^\alpha}_{rs}} L 
\end{eqnarray}
In the subsequent derivations only the $\alpha$-electron spin channel is
considered hence the $\alpha$ label is dropped. The resulting equations 
directly transfer to the $\beta$-electron spin channel as well.
For the orthogonality of the correlation functions we obtain from
$\partial L/\partial\lambda^C_{rs} = 0$
\begin{eqnarray}
  \sum_{i,j=1}^{n_b}C^*_{ir}I_{ij}C_{js} &=& I_{rs}
\end{eqnarray}
For the natural orbitals 
$\partial L/\partial\lambda^N_{ij} = 0$ gives
\begin{eqnarray}
  d_i\left(I_{ij}-\sum_{a,b=1}^{n_b}N^*_{ai}S_{ab}N_{bj}\right) &=& 0
\end{eqnarray}
which is also satisfied if the expression in brackets is zero leading to
\begin{eqnarray}
  \sum_{a,b=1}^{n_b}N^*_{ai}S_{ab}N_{bj} &=& I_{ij}
\end{eqnarray}
Furthermore, from $\partial L/\partial N^*_{ck} = 0$ we have for the
natural orbitals
\begin{eqnarray}
  \left(\sum_{b=1}^{n_b}F_{cb}N_{bk}
       -\sum_{b,j=1}^{n_b}S_{cb}N_{bj}\lambda^N_{jk}\right)
  d_k &=& 0
  \label{Eq:eig-natorb-d}
\end{eqnarray}
where $F$ is the Fock matrix. The Fock matrix is the matrix representation of
an effective one-electron operator which is derived from the total energy 
expression by differentiating it with respect to the density matrix.
More specifically the Fock matrix elements $F_{ab}$ are defined as
\begin{equation}
  F_{ab} = \frac{\partial E(D)}{\partial D_{ab}}
  \label{Eq:fock-atorb}
\end{equation}
Eq.~\ref{Eq:eig-natorb-d} is also satisfied if we solve
\begin{eqnarray}
  \sum_{b=1}^{n_b} F_{cb}N_{bk} &=& \sum_{b,j=1}^{n_b}S_{cb}N_{bj}\lambda^N_{jk}
  \label{Eq:eig-natorb}
\end{eqnarray}
instead. 
For the correlation functions we find from
$\partial L/\partial C^*_{kt} = 0$
\begin{eqnarray}
  0 &=& \sum_{i=1}^{n_b} F^{N'}_{ki}C_{ic}
  + \sum_{j=1}^{n_b}C_{kt}\lambda^N_{kj}
    \left(I_{kj}-\sum_{a,b=1}^{n_b}N^*_{ak}S_{ab}N_{bj}\right) 
  - \sum_{j,s=1}^{n_b}I_{kj}C_{js}\lambda^C_{st}
  \label{Eq:fock-corr-a} \\
  F^{N'}_{ki} &=& \sum_{a,b=1}^{n_b}N^*_{ak}F_{ab}N_{bk}\delta_{ki}
  \label{Eq:fock-corr}
\end{eqnarray}
where the second term in Eq.~\ref{Eq:fock-corr-a} is identical zero because of the
orthogonality of the
natural orbitals. Hence this equation simplifies to
\begin{eqnarray}
  \sum_{i=1}^{n_b} F^{N'}_{ki}C_{it}
  &=& \sum_{j,s=1}^{n_b}I_{kj}C_{js}\lambda^C_{st} 
  \label{Eq:eig-corr}
\end{eqnarray}
Therefore the wave function optimization problem translates into two secular
equations. The first one, Eq.~\ref{Eq:eig-natorb}, is a secular equation
for the natural orbitals and is in fact the same as the Kohn-Sham or
Hartree-Fock equation. The second one, Eq.~\ref{Eq:eig-corr}, seems very
similar to the first one except that it solves for the correlation functions.
One relevant difference is that the structure of the Fock matrix $F^{N'}_{ki}$
is special in that it is a diagonal matrix in the natural orbital basis. 
This may seem strange as it is counter intuitive that a non-trivial rotation
can be obtained by diagonalizing a matrix based on a matrix that is diagonal
already. To explain this situation consider the 2x2 problem 
\begin{eqnarray}
  \left(\begin{array}{cc}
  f_{11} & 0      \\
  0      & f_{22}
  \end{array}\right)
  \left(\begin{array}{cc}
  c_1 & -c_2   \\
  c_2 &  c_1
  \end{array}\right)
  &=& 
  \left(\begin{array}{cc}
  c_1 & -c_2   \\
  c_2 &  c_1
  \end{array}\right)
  \left(\begin{array}{cc}
  \lambda_1 & 0        \\
  0         & \lambda_2
  \end{array}\right)
\end{eqnarray}
There are two possible scenarios for the eigenvectors $(c_1,c_2)^T$ and
$(-c_2,c_1)^T$. A trivial result is obtained by choosing the eigenvectors to
be unit vectors, e.g. by choosing $c_1=1$ and $c_2=0$. In that case
$\lambda_1 = f_{11}$ and $\lambda_2 = f_{22}$. 
The other case, when the eigenvectors are not unit vectors, i.e. $c_1 \ne 0$
and $c_2 \ne 0$, requires
\begin{eqnarray}
  \left(\begin{array}{c}
  f_{11}c_1 \\
  f_{22}c_2
  \end{array}\right)
  &=& 
  \lambda_1
  \left(\begin{array}{c}
  c_1 \\
  c_2 
  \end{array}\right)
\end{eqnarray}
and therefore
that $\lambda_1 = f_{11} = f_{22}$, and likewise for $\lambda_2$. Therefore
this scenario has a solution only when the diagonal elements are the same, and
correspondingly, the eigenvalues are degenerate. As the diagonal elements are
the expectation values of the natural orbitals this implies that the natural
orbitals are degenerate as well. Such a solution clearly complies with the 
conditions set out in Gilbert's theorem for correlated orbitals. I.e. that at
convergence the energies of all orbitals with fractional occupation numbers
must be degenerate. Here this also translates into the corresponding generalized
orbitals being degenerate. Because every generalized orbital represents a 
single electron in a single-determinant wave function this also implies that 
all correlated electrons are degenerate.

Another corollary of these considerations is that starting a calculation
choosing
the correlation functions to be unit vectors is a particularly bad choice. 
Because these unit vectors are always a solution to Eq.~\ref{Eq:eig-corr}
there is no way to find a correlated state even if the energy expression 
accounts for electron correlation. Instead calculations have to be started 
assuming that initially the electrons are distributed over all relevant
natural orbitals. If the energy expression accounts for electron correlation
the expressions will converge towards degenerate correlation functions
that distribute the electrons over the natural orbitals. If the energy
expression has no electron correlation that favors fractionally occupied
natural orbitals the correlation functions will converge towards unit vectors.

Although it would seem that Eq.~\ref{Eq:eig-natorb} and~\ref{Eq:eig-corr}
can be solved straightforwardly in conventional matrix diagonalization based
ways the degeneracies cause complications. As at convergence all correlated
orbitals are degenerate we find that any arbitrary rotation among those orbitals
is an equally valid solution of the matrix diagonalization. In practice only
some of these solutions minimize the total energy but diagonalizing the Fock
operator cannot identify those solutions. Hence attempts to solve these
equations using a diagonalization based approach will very likely fail. 

Instead we construct gradients as
\begin{eqnarray}
   F^N_{ij} &=& \sum_{a,b=1}^{n_b} N^*_{ai}F_{ab} N_{bj} \\
   F^C_{rs} &=& \sum_{i,j=1}^{n_b} C^*_{ir}F^{N'}_{ij} C_{js}
\end{eqnarray}
and from these skew symmetric matrices are formed as
\begin{eqnarray}
   T^N &=& \left(\begin{array}{cccc}
           0          & -F^N_{12}  & \ldots & -F^N_{1n_b} \\
           F^N_{21}   & 0          & \ldots & -F^N_{2n_b} \\
           \vdots     & \vdots     & 0      & \vdots      \\
           F^N_{n_b1} & F^N_{n_b2} & \ldots & 0 \\
           \end{array}\right) \\
   T^C &=& \left(\begin{array}{cccc}
           0          & -F^C_{12}  & \ldots & -F^C_{1n_b} \\
           F^C_{21}   & 0          & \ldots & -F^C_{2n_b} \\
           \vdots     & \vdots     & 0      & \vdots      \\
           F^C_{n_b1} & F^C_{n_b2} & \ldots & 0 \\
           \end{array}\right)
\end{eqnarray}
Rotation matrices for the natural orbitals and correlation functions may be
written as
\begin{eqnarray}
   R^N &=& e^{\sigma_N T^N} \\
   R^C &=& e^{\sigma_C T^C}
\end{eqnarray}
The values of $\sigma_N$ and $\sigma_C$ are established by a line 
search~\cite{Nocedal20063} for the minimum of the
total energy. After updating the natural orbitals and correlation functions
a new line search is started. The iterations over line searches continue until
the optimal $\sigma$-s are zero. By explicitly choosing rotations that minimize
the total energy the indeterminateness of the eigenvectors of a degenerate
matrix is circumvented.
At present the line searches are performed alternately between $\sigma_N$ and
$\sigma_C$. 

\subsection{Equations of motion}
\label{sect:opt-wfn:eom}

In order for the expression proposed in Eq.~\ref{Eq:wavefunction} to be a 
proper wave function it must satisfy certain equations of motion. Of course
one can substitute this expression into the time dependent Schr{\"o}dinger
equation. As the expression is only a single-determinant wave function and
one that only generates the exact one-electron density matrix (without any
considerations to the accuracy of the corresponding two-electron density
matrix) there seems to be little to be gained over Hartree-Fock by doing that. 
Instead it seems wise to focus on the one thing the wave function can represent 
exactly, the one-electron density matrix, and consider its equation of motion
which is given by the von Neumann 
equation~\cite{Neumann1927} 
\begin{equation}
   i\hbar \frac{\mathrm{d}D}{\mathrm{d}t} = [H,D]
\end{equation}
In the case that the Hamiltonian is time independent the
time dependent density matrix is given by
\begin{equation}
   D(t) = e^{-iHt/\hbar}D(0)e^{iHt/\hbar}
\end{equation}
It would seem that the most straightforward way to evaluate the time dependent
density matrix is to make use of Eq.~\ref{Eq:genorb} allowing the generalized
orbital to be written as a vector. In addition the Hamiltonian can be
generalized to
\begin{equation}
   H = \left(
   \begin{array}{cccc}
      F      & 0      & \ldots & 0      \\
      0      & F      & \ldots & 0      \\
      \vdots & \vdots & \ddots & \vdots \\
      0      & 0      & \ldots & F 
   \end{array}
   \right)
   \label{Eq:genh}
\end{equation}
where $F$ is the effective one-electron Hamiltonian, the Fock matrix, of
Eq.~\ref{Eq:fock-atorb} replicated along the diagonal so that
\begin{equation}
   H_{ai,bj} = \delta_{ij}F_{ab}
\end{equation}
This approach is similar to the approach applied to the metric $S^G$ but here
the block diagonal representation is favored over the form of
Eq.~\ref{Eq:metric}. The reason is that the Fock matrix of Eq.~\ref{Eq:genh}
is to be used specfically with a orthonormal set of generalized orbitals
associated with a particular single-determinantal wave function. The form of
the metric in Eq.~\ref{Eq:metric} on the contrary has been generalized to allow
calculating the overlap between generalized orbitals of different 
single-determinant wave functions.

Subsequently evaluating 
\begin{equation}
   G_{ai,r}(t) = \sum_{b,j=1}^{n_b}e^{-iHt/\hbar}G_{bj,r}(0)
\end{equation}
the generalized orbitals at time $t$ are obtained from which $D(t)$ can be
constructed according to
\begin{equation}
   D_{ab}(t) = \sum_{i=1}^{n_b}\sum_{r=1}^{n_e}G_{ai,r}(t)G^*_{bi,r}(t)
\end{equation}
This equation is essentially the same as Eq.~\ref{Eq:densmat} except that nor
the natural orbitals nor the correlation functions are explicitly referenced.
In order to return the generalized orbitals to the form of Eq.~\ref{Eq:genorb}
the density matrix $D(t)$ can be diagonalized to obtain the natural orbitals
at time $t$. Furthermore time dependent correlation functions can be 
optimized that generate the occupation numbers at time $t$.

The approach to optimizing the correlation functions simply involves writing
the correlation functions at time $t$ in
terms of the correlation functions at time $0$ as
\begin{equation}
   C_{ir}(t) = \sum_{s=1}^{n_b}C_{is}(0)U_{sr}(t)
\end{equation}
where $U$ is a unitary matrix. Subsequently the error 
\begin{eqnarray}
   R &=& \sum_{i=1}^{n_b}
         \left(d'_i(t)-d_i(t)\right)^2
      +  \sum_{r,s=1}^{n_b}\lambda_{rs}
         \left(\delta_{rs}-\sum_{i=1}^{n_b}C_{ir}(t)C^*_{is}(t)\right) \\
   d'_i(t) &=& \sum_{r=1}^{n_e}C_{ir}(t)C_{ir}^*(t)
\end{eqnarray}
as a function of $U$ is minimized to obtain the eigenvalue equation
\begin{eqnarray}
   2\sum_{i,s=1}^{n_b}C^*_{ir}(0)(d'_i(t)-d_i(t))C_{is}(0)U_{st}
   &=& \sum_{s=1}^{n_b}U_{rs}\lambda_{st}
\end{eqnarray}
where $d_i(t)$ are the occupation numbers obtained from the diagonalization of
$D(t)$.
This equation needs to be solved iteratively as $d'_i(t)$ depends on the
solution.

Supposedly this approach can also be extended to time dependent Hamiltonians 
by introducing sufficiently small time steps.
Hence it would seem that the equations of motion as given by the von Neumann
equation can be applied relatively straightforwardly in the present context.
Admittedly to date we have not tested this approach as yet.

\section{Wavefunction analysis}
\label{sect:wfn-ana}

Once the wave function has been converged we have a set of orbitals and the
total
energy of the system. Moreover, if there are degenerate energy levels with
fractional 
occupation numbers in the natural orbitals the corresponding orbital energies 
equal the chemical potential. For the orbitals obtained we know that they 
generate the one-electron density matrix that minimizes the total energy. In
practice the corresponding set of orbitals is not unique. Like in Hartree-Fock
or Kohn-Sham theory the total energy is invariant under some rotations among
the orbitals. Furthermore, in Hartree-Fock theory Koopmans
theorem~\cite{Koopmans1934} can be used
to estimate ionization potentials and electron affinities from the orbital
energies. The chemical potential obtained at the end of the optimization does
not afford the same rich set of information. In this section we will
show that in our approach some of the information resulting from a Hartree-Fock 
calculation can be obtained from our wave function as well with modest extra
effort.

The first important point to note is that as the energy is a functional of the
one-electron density matrix any transformations of the wave function that leave
this matrix unchanged do not affect the total energy either. In practice this
means that rotations among natural orbitals with equal occupation numbers
leave the total energy unchanged. Likewise rotations amongst the occupied 
correlation functions, or rotations amongst the unoccupied correlation functions
do not affect the total energy. While the total energy for the ground state
remains unchanged under these rotations processes that change the orbital 
occupations might be very sensitive to the particulars of the orbitals. An 
example of such a process is the excitation of an electron.
Hence, specific orbital representations that support meaningful interpretations 
are highly desirable.

In practice the natural orbitals cannot be canonicalized. For the natural
orbitals the ordering is critical to maintain the link with the correlation
functions. This fact in combination with the degeneracies cause any approach
to canonicalization to fail on these orbitals.

The correlation functions can be canonicalized in a more informative way. To 
this end the Fock matrix of the traditional Hartree-Fock energy is evaluated.
The expectation values of the natural orbitals over this matrix feed into the
Fock matrix for the correlation functions. This Fock matrix is transformed into
the correlation function basis and the occupied-unoccupied block zeroed to 
suppress mixings that change the density matrix. The resulting matrix is
diagonalized to obtain canonical correlation functions. Because the Fock
matrix of the
Hartree-Fock energy has few degenerate states the resulting orbitals and their
energies are Hartree-Fock like. Nevertheless, these orbitals still generate the
same correlated one-electron density matrix. This combination suggests that the
orbitals can be interpreted similarly to the ones from Hartree-Fock theory. 
This option was obtained at the cost of a single Fock matrix construction which
is modest compared to the few dozen Fock matrix constructions that are required
in the self consistent field procedure. Obviously the correlation functions may
be canonicalized over any operator. Which operator to choose depends on what the
results are to be used for.

\section{Application to \ce{LiH} and \ce{Be}}
\label{sect:app}

In order to demonstrate what the formalism introduced can provide it is applied
to the systems \ce{LiH} and \ce{Be}. The reasons for this choice are that both
systems are small so the results can be reported in detail. The \ce{Be} atom is
a system
with well known near degenerate states whereas the \ce{LiH} molecule is a weakly
correlated
system. The energy expression used is the power functional~\cite{Muller1984}
with a power of 0.578~\cite{Lathiotakis2009}. The formalism and the 
functional were implemented in a modified version of NWChem~\cite{Valiev2010}.
The electronic structure was described using the 6-31G basis
set~\cite{Hehre1972,Dill1975,Schuchardt2007}. In \ce{LiH} the bond length
was 1.5957 \AA.

The total energies for \ce{Be} and \ce{LiH} are reported in 
Tables~\ref{table:be-tot-en} and~\ref{table:lih-tot-en}, respectively.
The approach proposed in this work is referred to as Wave Function based Density
Matrix Functional Theory (WFDMFT). The occupation numbers for \ce{Be} and
\ce{LiH} are compared with those of the Full-CI method in 
Tables~\ref{table:be-tot-occ} and~\ref{table:lih-tot-occ}. The orbital energies
are given in Tables~\ref{table:be-tot-orb} and~\ref{table:lih-tot-orb}.
The results under the columns "converged" are the orbital energies at the
end of the power functional optimization. The results under the column 
"canonicalized" are the orbital energies obtained from the canonicalization of
the correlation functions over the Fock matrix from Hartree-Fock.
Finally, Tables~\ref{table:be-exc-en} and~\ref{table:lih-exc-en} give the
total energies obtained from excited Slater determinants of generalized
orbitals. The full wave functions for \ce{Be} and \ce{LiH} are given in the
Supplemental Material~\cite{vanDamsupplement}.

The total energy obtained from the WFDMFT method as well as the fractional
occupation numbers for the natural orbitals show that correlation is
accounted for to some degree. Obviously this is only to be expected from a
reduced density matrix functional theory. More interesting is that the orbital
picture behind the WFDMFT method enables multiple perspectives on the results.
The orbital energies from the power functional display the degeneracies that
Gilbert's theorem calls for and reveal the chemical potential. Yet when the
correlation functions are canonicalized over the regular Fock operator from
Hartree-Fock energy levels similar to those of Hartree-Fock theory are
recovered.
Furthermore, moving one electron from one occupied correlation function to
an unoccupied one a Slater determinant of an excited state is obtained. These
latter features are trivial in this method but are cumbersome in other
reduced density matrix functional approaches.

\begin{table}
\caption{The total energies ($/E_H$) for \ce{Be} in the 6-31G basis set for
         different methods}
\label{table:be-tot-en}
\begin{tabular}{cc}
\hline
Hartree-Fock & -14.566764 \\
WFDMFT       & -14.590417 \\
Full-CI      & -14.613545 \\
\hline
\end{tabular}
\end{table}

\begin{table}
\caption{The total energies ($/E_H$) for \ce{LiH} in the 6-31G basis set for
         different methods}
\label{table:lih-tot-en}
\begin{tabular}{cc}
\hline
Hartree-Fock & -7.979277 \\
WFDMFT       & -7.985189 \\
Full-CI      & -7.998288 \\
\hline
\end{tabular}
\end{table}

\begin{table}
\caption{Occupation numbers of the total one-electron density matrix for
         \ce{Be}}
\label{table:be-tot-occ}
\begin{tabular}{ccc}
\hline
Hartree-Fock & WFDMFT & Full-CI \\
\hline
2.0000       & 2.0000 & 1.9999 \\
2.0000       & 1.8502 & 1.8010 \\
0.0000       & 0.0456 & 0.0657 \\
0.0000       & 0.0457 & 0.0657 \\
0.0000       & 0.0459 & 0.0657 \\
0.0000       & 0.0042 & 0.0020 \\
0.0000       & 0.0028 & 0.0000 \\
0.0000       & 0.0028 & 0.0000 \\
0.0000       & 0.0028 & 0.0000 \\
\hline
\end{tabular}
\end{table}

\begin{table}
\caption{Occupation numbers of the total one-electron density matrix for
         \ce{LiH}}
\label{table:lih-tot-occ}
\begin{tabular}{ccc}
\hline
Hartree-Fock & WFDMFT & Full-CI \\
\hline
2.0000       & 2.0000 & 1.9999 \\
2.0000       & 1.9501 & 1.9564 \\
0.0000       & 0.0297 & 0.0394 \\
0.0000       & 0.0023 & 0.0016 \\
0.0000       & 0.0023 & 0.0011 \\
0.0000       & 0.0045 & 0.0011 \\
0.0000       & 0.0009 & 0.0005 \\
0.0000       & 0.0018 & 0.0000 \\
0.0000       & 0.0018 & 0.0000 \\
0.0000       & 0.0051 & 0.0000 \\
0.0000       & 0.0013 & 0.0000 \\
\hline
\end{tabular}
\end{table}

\begin{table}
\caption{Orbital energies of \ce{Be}}
\label{table:be-tot-orb}
\begin{tabular}{ccc}
\hline
Hartree-Fock & \multicolumn{2}{c}{WFDMFT} \\
             & converged & canonicalized \\
\hline
-4.7069      &   -3.7562 & -4.7221 \\
-0.3013      &   -0.1589 & -0.2565 \\
 0.0824      &   -0.1589 &  0.0506 \\
 0.0824      &   -0.1589 &  0.0757 \\
 0.0824      &   -0.1589 &  0.0758 \\
 0.4398      &   -0.1589 &  0.4351 \\
 0.4649      &   -0.1589 &  0.4543 \\
 0.4649      &   -0.1589 &  0.4570 \\
 0.4649      &   -0.1589 &  0.4570 \\
\hline
\end{tabular}
\end{table}

\begin{table}
\caption{Orbital energies of \ce{LiH}}
\label{table:lih-tot-orb}
\begin{tabular}{ccc}
\hline
Hartree-Fock & \multicolumn{2}{c}{WFDMFT} \\
             & converged & canonicalized \\
\hline
-2.4533      & -1.7574   & -2.4561 \\
-0.3007      & -0.1068   & -0.2893 \\
 0.0094      & -0.1068   &  0.0033 \\
 0.0602      & -0.1068   &  0.0588 \\
 0.0602      & -0.1068   &  0.0596 \\
 0.1449      & -0.1068   &  0.1413 \\
 0.2005      & -0.1068   &  0.2005 \\
 0.2219      & -0.1068   &  0.2196 \\
 0.2219      & -0.1068   &  0.2205 \\
 0.3598      & -0.1068   &  0.3546 \\
 1.3182      & -0.1068   &  1.3091 \\
\hline
\end{tabular}
\end{table}

\begin{table}
\caption{Total energies of excited determinants of \ce{Be}}
\label{table:be-exc-en}
\begin{tabular}{ccc}
\hline
Excitation              & Hartree-Fock & WFDMFT \\
\hline
ground state            & -14.566764   & -14.590417 \\
$\alpha:2\rightarrow 3$ & -14.425466   & -14.475988 \\
$\alpha:2\rightarrow 4$ & -14.425466   & -14.444019 \\
$\alpha:2\rightarrow 5$ & -14.425466   & -14.444326 \\
\hline
\end{tabular}
\end{table}

\begin{table}
\caption{Total energies of excited determinants of \ce{LiH}}
\label{table:lih-exc-en}
\begin{tabular}{ccc}
\hline
Excitation              & Hartree-Fock & Total Energy \\
\hline
ground state            & -7.979277    & -7.985189  \\
$\alpha:2\rightarrow 3$ & -7.829456    & -7.835788  \\
$\alpha:2\rightarrow 4$ & -7.789693    & -7.794577  \\
$\alpha:2\rightarrow 5$ & -7.789693    & -7.792494  \\
\hline
\end{tabular}
\end{table}

\section{Conclusions}

This papers demonstrates that a fictitious one-electron system can be 
formulated that can represent any arbitrary one-electron density matrix 
with a single-determinant wave function. This includes even density matrices
corresponding to a correlated state. The formulation of the wave function
requires the conventional natural orbitals as well as a new set referred to
as correlation functions. Optimizing such a wave function requires solving
the traditional Hartree-Fock or Kohn-Sham equations, with a suitable
reduced density matrix functional, as well as a new set of equations for
the correlation functions. Both sets of equations are very similar in 
structure. 
The analysis of the wave function allows for representations that are very
similar to the orbital energies of Hartree-Fock theory. In addition excited
determinants can easily be generated by applying the usual excitation 
operators to the ground state determinant.

\begin{acknowledgments}
This manuscript has been authored by employees of Brookhaven Science Associates,
LLC under Contract No. DE- SC0012704 with the U.S. Department of Energy.

A portion of the research was performed using EMSL, a DOE Office of Science
User Facility sponsored by the Office of Biological and Environmental Research
and located at Pacific Northwest National Laboratory.
\end{acknowledgments}


\bibliographystyle{unsrt}
\bibliography{RDMFT_KS,RDMFT_KS_alt}

\end{document}


\title{Generalization of the Kohn-Sham system that can represent arbitary
       one electron density matrices \\ Supplemental material}
\author{Hubertus J. J. van Dam}
\affiliation{Brookhaven National Laboratory, Upton, NY 11973-5000}
\date{April 26, 2016}

\begin{abstract}
This document provides an illustrative example of the key aspects that 
enable the formulation of a single determinant wave functon that generate
an arbitrary one-electron density matrix.
It also provides the converged and canonicalized wave functions of 
\ce{Be} and \ce{LiH}. Readers are referred to the main article for details.
\end{abstract}

\maketitle

\section{Illustration of the key aspects of the wave function definition}

In section~\ref{P-sect:def-wfn} of the main article a wave function was defined
to generate the exact
one-electron density matrix. The approach involves mapping an $n_b$ dimensional
problem into an $n_b^2$ dimensional space and introducing a projection that 
takes an $n_b^2$ dimensional problem back to an $n_b$ dimensional one. To clarify
the key points we will briefly consider a 2-dimensional example.

When optimizing a one-electron density matrix transformations that change the
natural orbitals as well as the occupation numbers have to be considered. Let
us consider transforming a one-electron density $D$ into a one-electron
density matrix $D'$ both represented in a 2-dimensional basis. The matrix 
$D$ and $D'$ are given by
\begin{eqnarray}
   D  &=& \left(\begin{array}{cc}
          1 & 0 \\
          0 & 0 
          \end{array}\right)
   \label{Eq:D} \\
   D' &=& \left(\begin{array}{cc}
          1/2 & 0 \\
          0   &  1/2
          \end{array}\right)
   \label{Eq:Dprime}
\end{eqnarray}
Both density matrices clearly satisfy the N-representability conditions 
of the one-electron density matrices of a single spin channel containing one
electron. 
The most general approach to transform $D$ into $D'$ involves different left and
right transformations, i.e. a transformation of the form
\begin{eqnarray}
   D' &=& L^T D R \\
   D'_{il} &=& \sum_{j,k=1}^{2} L_{ji}D_{jk}R_{kl}
\end{eqnarray}
From Eqs.~\ref{Eq:D} and~\ref{Eq:Dprime} four equations that $L$ and $R$ need
to satisfy are obtained to generate the elements of $D'$
\begin{eqnarray}
   D'_{11}: && 1/2 = L_{11}R_{11} \label{Eq:Dprime_11}\\
   D'_{21}: && 0   = L_{12}R_{11} \label{Eq:Dprime_21}\\
   D'_{12}: && 0   = L_{11}R_{12} \label{Eq:Dprime_12}\\
   D'_{22}: && 1/2 = L_{12}R_{12} \label{Eq:Dprime_22}
\end{eqnarray}
Clearly Eqs.~\ref{Eq:Dprime_11} and~\ref{Eq:Dprime_22} require that $L_{11}$,
$R_{11}$, $L_{12}$ and $R_{12}$ all be non-zero, but Eqs.~\ref{Eq:Dprime_21} 
and~\ref{Eq:Dprime_12} require either $L_{12}$ or $R_{11}$ to be zero and 
either $L_{11}$ or $R_{12}$ to be zero. Obviously these equations represent
a contradiction and in a 2-dimensional space there cannot exist a transformation
that transforms $D$ into $D'$.

The approach taken in section~\ref{P-sect:def-wfn} of the main article is to map
this problem into a $n_b^2$ 
dimensional space or in this example a 4-dimensional space. For this example the natural
orbitals of both matrices $D$ and $D'$ are the same and are
\begin{eqnarray}
   N_1 &=& \left(\begin{array}{c}
           1 \\
           0
           \end{array}\right) \\
   N_2 &=& \left(\begin{array}{c}
           0 \\
           1
           \end{array}\right)
\end{eqnarray}
For this simple example there is only one occupied orbital and the corresponding
correlation function for $D$ is
\begin{eqnarray}
   C_1 &=& \left(\begin{array}{c}
           1 \\
           0
           \end{array}\right) \\
\end{eqnarray}
whereas the correlation function for $D'$ is
\begin{eqnarray}
   C'_1 &=& \left(\begin{array}{c}
            \sqrt{1/2} \\
            \sqrt{1/2}
            \end{array}\right) \\
\end{eqnarray}
Subsequently density matrices in $n_b^2$-dimensional space can be formulated by 
applying the routine definition of the density matrix of
Eq.~\ref{P-Eq:densmat_occ} of the main article
to the generalized orbitals rather than the natural orbitals. If we denote the 
4-dimensional density matrices $\bar{D}$ and $\bar{D}'$ they are given by
\begin{eqnarray}
   \bar{D}  &=& \left(\begin{array}{cccc}
                1 & 0 & 0 & 0 \\
                0 & 0 & 0 & 0 \\
                0 & 0 & 0 & 0 \\
                0 & 0 & 0 & 0 
                \end{array}\right)
   \label{Eq:Db} \\
   \bar{D}' &=& \left(\begin{array}{cccc}
          1/2 & 0 & 0 & 1/2 \\
          0   & 0 & 0 & 0   \\
          0   & 0 & 0 & 0   \\
          1/2 & 0 & 0 & 1/2 \\
          \end{array}\right)
   \label{Eq:Dbprime}
\end{eqnarray}
The matrix $\bar{D}'$ can easily be obtained from $\bar{D}$ using a rotation $\bar{R}$
through
\begin{eqnarray}
  \bar{D}' &=& \bar{R}^T\bar{D}\bar{R} \\
  \bar{R}  &=& \left(\begin{array}{cccc}
               \sqrt{1/2} & 0 & 0 & \sqrt{1/2} \\
               0          & 1 & 0 & 0          \\
               0          & 0 & 1 & 0          \\
              -\sqrt{1/2} & 0 & 0 & \sqrt{1/2} 
               \end{array}\right)
\end{eqnarray}
Finally the projection $p: \bar{D} \rightarrow D$ and $p: \bar{D}' \rightarrow D'$ 
as given in Eq.~\ref{P-Eq:densmat} of the main article is effected by summing the $n_b \times n_b$ diagonal
blocks of $\bar{D}$ and $\bar{D}'$ respectively. I.e.
\begin{eqnarray}
   D_{ab}  &=& \bar{D}_{ab}  + \bar{D}_{(a+2)(b+2)}, \;\;
   a,b \in \{1,2\} \\
           &=& \left(\begin{array}{cc}
               1 & 0 \\
               0 & 0 
               \end{array}\right) + 
               \left(\begin{array}{cc}
               0 & 0 \\
               0 & 0 
               \end{array}\right)  \\
   D'_{ab} &=& \bar{D}'_{ab} + \bar{D}'_{(a+2)(b+2)}, \;\;
   a,b     \in \{1,2\} \\
           &=& \left(\begin{array}{cc}
               1/2 & 0 \\
               0 & 0 
               \end{array}\right) + 
               \left(\begin{array}{cc}
               0 & 0 \\
               0 & 1/2
               \end{array}\right) 
\end{eqnarray}

Hence it has been demonstrated that density matrix transformations that cannot
be represented in an $n_b$-dimensional space can be effected using traditional
rotation operations by mapping the density matrix into an $n_b^2$-dimensional 
space, applying the transformation and projecting the result back to the
original $n_b$-dimensional space. This approach enables the calculation of
a non-idem-potent one-electron density matrix from a single determinant
wave function.

Formally, the decomposition of the generalized orbitals into natural orbitals and
correlation functions is not required. This decomposition is preferred purely to
keep the
numerical complexity limited to $O(n_b^3)$. Otherwise the transformations of
$n_b^2$-dimensional  matrices would have a complexity of $O(n_b^6)$. Even
manipulating 
generalized orbitals of which there are $n_b$ linearly independent ones with 
$n_b^2 \times n_b^2$ transformation matrices would still have a complexity of
$O(n_b^5)$. Hence decomposing the generalized orbitals is well justified on
the grounds of computational cost but formally not required.

\section{The WFDMFT wave function for \ce{Be} in the 6-31G basis set}

The wave function reported here was obtained by minimizing the energy of the
power functional~\cite{Muller1984} with a power of
0.578~\cite{Lathiotakis2009} for \ce{Be}. The wave function is expressed in the 6-31G
basis set~\cite{Hehre1972,Dill1975,Schuchardt2007}. The calculations were
performed with a modified version of NWChem~\cite{Valiev2010}. As the ground
state of \ce{Be} is a singlet state the $\alpha$- and $\beta$-electron
wave functions are the same. Hence only the $\alpha$-electron wave function is
reported.

The basis is expressed in terms of linear combinations of Cartesian Gaussian
basis functions centered at the position of the \ce{Be} nucleus. The ordering
of the basis functions is given in Table~\ref{tab:be_bas_order} and the
exponents and contraction coefficients are given in Table~\ref{tab:be_bas}.
The natural orbitals are given in Table~\ref{tab:be_nat_orb} and the 
correlation functions are given in Table~\ref{tab:be_corr_fun}. 

\begin{table}
\caption{The ordering of the basis functions for the 6-31G basis set of \ce{Be}}
\label{tab:be_bas_order}
\begin{tabular}{cccccccccc}
\hline
Number   & 1    & 2    & 3      & 4      & 5      & 6    & 7      & 8      & 9      \\
Function & $1S$ & $2S$ & $2P_x$ & $2P_y$ & $2P_z$ & $3S$ & $3P_x$ & $3P_y$ & $3P_z$ \\
\hline
\end{tabular}
\end{table}

\begin{table}
\caption{The exponents and contraction coefficients of the radial parts of the
         6-31G basis functions for \ce{Be}}
\label{tab:be_bas}
\begin{tabular}{crrr}
\hline
Number & Exponents    & $S$-coefficients & $P$-coefficients \\
\hline
1      & 1264.5857000 &  0.0019448       &                  \\
       &  189.9368100 &  0.0148351       &                  \\
       &   43.1590890 &  0.0720906       &                  \\
       &   12.0986630 &  0.2371542       &                  \\
       &    3.8063232 &  0.4691987       &                  \\
       &    1.2728903 &  0.3565202       &                  \\
\hline
2      &    3.1964631 & -0.1126487       &  0.0559802       \\
       &    0.7478133 & -0.2295064       &  0.2615506       \\
       &    0.2199663 &  1.1869167       &  0.7939723       \\
\hline
3      &    0.0823099 &  1.0000000       &  1.0000000       \\
\hline
\end{tabular}
\end{table}

\begin{table}
\caption{The natural orbitals of \ce{Be}}
\label{tab:be_nat_orb}
\begin{tabular}{crrrrrrrrr}
\hline
occupation &  1.0000000 &  0.9251160 &  0.0228225 &  0.0228689 &  0.0229336 &  0.0020900 &  0.0013902 &  0.0013865 &  0.0013924 \\
\hline
number     &          1 &          2 &          3 &          4 &          5 &          6 &          7 &          8 &          9 \\
    1      &  0.9967726 & -0.2280916 & -0.0000000 & -0.0000000 & -0.0000000 &  0.0018142 &  0.0000000 &  0.0000000 & -0.0000000 \\
    2      &  0.0195829 &  0.2829467 & -0.0000000 &  0.0000000 &  0.0000000 &  2.0180390 &  0.0000000 &  0.0000000 & -0.0000000 \\
    3      & -0.0000000 & -0.0000000 &  0.3264994 & -0.0568546 & -0.0191530 &  0.0000000 & -0.6627271 & -0.1537898 &  1.1124474 \\
    4      &  0.0000000 &  0.0000000 &  0.0587704 &  0.2891191 &  0.1547850 & -0.0000000 &  1.1015539 &  0.1602899 &  0.6785889 \\
    5      &  0.0000000 & -0.0000000 & -0.0098803 & -0.1552349 &  0.2960959 &  0.0000000 & -0.2169876 &  1.2842549 &  0.0483375 \\
    6      & -0.0031030 &  0.7663844 &  0.0000000 & -0.0000000 & -0.0000000 & -1.8961496 & -0.0000000 & -0.0000000 &  0.0000000 \\
    7      &  0.0000000 &  0.0000000 &  0.7348388 & -0.1272201 & -0.0423495 & -0.0000000 &  0.5690172 &  0.1321133 & -0.9546716 \\
    8      & -0.0000000 & -0.0000000 &  0.1326176 &  0.6479386 &  0.3449804 &  0.0000000 & -0.9463189 & -0.1381550 & -0.5827258 \\
    9      &  0.0000000 &  0.0000000 & -0.0219757 & -0.3472333 &  0.6585646 & -0.0000000 &  0.1861575 & -1.1042838 & -0.0417182 \\
\hline
\end{tabular}
\end{table}

\begin{table}
\caption{The canonical correlation functions of \ce{Be}}
\label{tab:be_corr_fun}
\begin{tabular}{crrrrrrrrr}
\hline
occupation   &  1.000000  &  1.000000  &  0.000000  &  0.000000  &  0.000000  &  0.000000  &  0.000000  &  0.000000  &  0.000000  \\
$E_{WFDMFT}$ & -3.756184  & -0.158908  & -0.158909  & -0.158909  & -0.158909  & -0.158909  & -0.158909  & -0.158909  & -0.158909  \\
$E_{HF}$     & -4.722078  & -0.256489  &  0.050647  &  0.075744  &  0.075768  &  0.435063  &  0.454294  &  0.456964  &  0.457005  \\
\hline
number       &          1 &          2 &          3 &          4 &          5 &          6 &          7 &          8 &          9 \\
    1        & -1.0000000 &  0.0000000 & -0.0000000 &  0.0000000 &  0.0000000 & -0.0000000 & -0.0000001 &  0.0000000 & -0.0000000 \\
    2        &  0.0000000 &  0.9618295 &  0.2641453 &  0.0001025 &  0.0001421 & -0.0471731 & -0.0537182 & -0.0004874 &  0.0003108 \\
    3        &  0.0000000 &  0.1510713 & -0.5565424 &  0.7764688 &  0.2530634 & -0.0148666 & -0.0164979 & -0.0001492 &  0.0000951 \\
    4        &  0.0000000 &  0.1512246 & -0.5567276 & -0.6070636 &  0.5460448 & -0.0148824 & -0.0165154 & -0.0001493 &  0.0000952 \\
    5        & -0.0000000 &  0.1514385 & -0.5569620 & -0.1690267 & -0.7986200 & -0.0149045 & -0.0165398 & -0.0001496 &  0.0000954 \\
    6        &  0.0000000 &  0.0457163 & -0.0109457 & -0.0000049 & -0.0000068 &  0.9931074 & -0.1073632 & -0.0008489 &  0.0005403 \\
    7        & -0.0000001 &  0.0372847 & -0.0084810 & -0.0000038 & -0.0000052 &  0.0601267 &  0.5717163 &  0.5411234 &  0.6125754 \\
    8        & -0.0000001 &  0.0372353 & -0.0084707 & -0.0000038 & -0.0000052 &  0.0601666 &  0.5803304 & -0.7965577 &  0.1537343 \\
    9        & -0.0000001 &  0.0373149 & -0.0084874 & -0.0000038 & -0.0000052 &  0.0601038 &  0.5666798 &  0.2695929 & -0.7753172 \\
\hline
\end{tabular}
\end{table}

\section{The WFDMFT wave function for \ce{LiH} in the 6-31G basis set}

The wave function reported here was obtained by minimizing the energy of the
power functional~\cite{Muller1984} with a power of
0.578~\cite{Lathiotakis2009} for \ce{LiH}. The wave function is expressed in the 6-31G
basis set~\cite{Hehre1972,Dill1975,Schuchardt2007}. The calculations were
performed with a modified version of NWChem~\cite{Valiev2010}. As the ground
state of \ce{LiH} is a singlet state the $\alpha$- and $\beta$-electron
wave functions are the same. Hence only the $\alpha$-electron wave function is
reported.

The basis is expressed in terms of linear combinations of Cartesian Gaussian
basis functions centered at the position of the \ce{Li} or \ce{H} nucleus. 
The geometry of the \ce{LiH} molecule is given in Table~\ref{tab:lih_geom}.
The ordering
of the basis functions is given in Table~\ref{tab:lih_bas_order} and the
exponents and contraction coefficients are given in Table~\ref{tab:lih_bas}.
The natural orbitals are given in Table~\ref{tab:lih_nat_orb} and the 
correlation functions are given in Table~\ref{tab:lih_corr_fun}. 

\begin{table}
\caption{The nuclear positions of \ce{LiH} in {\AA}ngstrom}
\label{tab:lih_geom}
\begin{tabular}{crrr}
\hline
Element & $x$    & $y$    & $z$    \\
\hline
Li      & 0.0000 & 0.0000 & 0.0000 \\
H       & 0.0000 & 0.0000 & 1.5957 \\
\hline
\end{tabular}
\end{table}

\begin{table}
\caption{The ordering of the basis functions for the 6-31G basis set of \ce{LiH}}
\label{tab:lih_bas_order}
\begin{tabular}{c||c|c|c|c|c|c|c|c|c|c|c}
\hline
Number   & 1            & 2            & 3              & 4              & 5              & 6            & 7              & 8              & 9              & 10          & 11          \\
Function & \ce{Li} $1S$ & \ce{Li} $2S$ & \ce{Li} $2P_x$ & \ce{Li} $2P_y$ & \ce{Li} $2P_z$ & \ce{Li} $3S$ & \ce{Li} $3P_x$ & \ce{Li} $3P_y$ & \ce{Li} $3P_z$ & \ce{H} $1S$ & \ce{H} $2S$ \\
\hline
\end{tabular}
\end{table}

\begin{table}
\caption{The exponents and contraction coefficients of the radial parts of the
         6-31G basis functions for \ce{Li} and \ce{H}}
\label{tab:lih_bas}
\begin{tabular}{crrr}
\hline
Number    & Exponents    & $S$-coefficients & $P$-coefficients \\
\hline
\ce{Li} 1 & 642.4189200  &       0.00214260 &                  \\
          &  96.7985150  &       0.01620890 &                  \\
          &  22.0911210  &       0.07731560 &                  \\
          &   6.2010703  &       0.24578600 &                  \\
          &   1.9351177  &       0.47018900 &                  \\
          &   0.6367358  &       0.34547080 &                  \\
\hline
\ce{Li} 2 &   2.3249184  &      -0.03509170 &        0.0089415 \\
          &   0.6324306  &      -0.19123280 &        0.1410095 \\
          &   0.0790534  &       1.08398780 &        0.9453637 \\
\hline
\ce{Li} 3 &   0.0359620  &       1.00000000 &        1.0000000 \\
\hline
\ce{H}  1 &  18.7311370  &       0.03349460 &                  \\
          &   2.8253937  &       0.23472695 &                  \\
          &   0.6401217  &       0.81375733 &                  \\
\hline
\ce{H}  2 &   0.1612778  &       1.00000000 &                  \\
\hline
\end{tabular}
\end{table}

\begin{sidewaystable}
\caption{The natural orbitals of \ce{LiH}}
\label{tab:lih_nat_orb}
\begin{tabular}{crrrrrrrrrrr}
\hline
occupation &  1.0000000 &  0.9750439 &  0.0148714 &  0.0011666 &  0.0011668 &  0.0022720 &  0.0004384 &  0.0009026 &  0.0009026 &  0.0025726 &  0.0006633 \\
\hline
number     &          1 &          2 &          3 &          4 &          5 &          6 &          7 &          8 &          9 &         10 &         11 \\
    1      &  0.9986362 & -0.1309242 & -0.1393341 & -0.0000000 & -0.0000000 & -0.0450261 &  0.0403052 & -0.0000000 & -0.0000000 &  0.0817203 & -0.1186674 \\
    2      &  0.0168104 &  0.2198689 & -0.1068899 &  0.0000000 &  0.0000000 & -0.0952209 &  1.9183083 & -0.0000000 & -0.0000000 &  1.6708956 & -0.3415826 \\
    3      &  0.0000000 & -0.0000000 & -0.0000000 &  0.0362089 &  0.0438189 &  0.0000000 &  0.0000000 &  0.4682339 &  1.6065657 &  0.0000000 & -0.0000000 \\
    4      & -0.0000000 &  0.0000000 &  0.0000000 &  0.0437525 & -0.0362411 & -0.0000000 & -0.0000000 &  1.6065674 & -0.4682320 & -0.0000000 &  0.0000000 \\
    5      & -0.0089540 &  0.2250094 & -0.2074908 &  0.0000000 & -0.0000000 & -0.6337137 & -0.9393584 &  0.0000000 &  0.0000000 &  1.7813064 & -0.3368418 \\
    6      & -0.0086715 &  0.1362832 &  0.8652146 & -0.0000000 &  0.0000000 &  0.7317441 & -1.9792789 &  0.0000000 &  0.0000000 & -0.6537602 & -0.2521012 \\
    7      & -0.0000000 &  0.0000000 &  0.0000000 &  0.6080391 &  0.7349034 & -0.0000000 & -0.0000000 & -0.3850335 & -1.3211671 & -0.0000000 &  0.0000000 \\
    8      &  0.0000000 &  0.0000000 & -0.0000000 &  0.7349580 & -0.6080126 &  0.0000000 &  0.0000000 & -1.3211395 &  0.3850653 &  0.0000000 & -0.0000000 \\
    9      &  0.0051263 & -0.0093754 & -0.4629523 & -0.0000000 &  0.0000000 &  1.3709321 &  0.5308445 & -0.0000000 & -0.0000000 & -0.6543060 & -0.2154257 \\
   10      &  0.0028701 &  0.2980824 & -0.0600038 &  0.0000000 & -0.0000000 & -0.0934628 & -0.0104652 &  0.0000000 &  0.0000000 & -0.0818974 & -1.3961472 \\
   11      &  0.0028801 &  0.4419011 & -0.0759035 &  0.0000000 & -0.0000000 & -0.5682003 &  0.2974165 & -0.0000000 & -0.0000000 & -1.4769395 &  1.7949134 \\
\hline
\end{tabular}
\end{sidewaystable}

\begin{sidewaystable}
\caption{The canonical correlation functions of \ce{LiH}}
\label{tab:lih_corr_fun}
\begin{tabular}{crrrrrrrrrrr}
\hline
occupation   &  1.000000  &  1.000000  &  0.000000  &  0.000000  &  0.000000  &  0.000000  &  0.000000  &  0.000000  &  0.000000  &  0.000000  &  0.000000  \\
$E_{WFDMFT}$ & -1.757418  & -0.106771  & -0.106778  & -0.106788  & -0.106777  & -0.106776  & -0.106773  & -0.106770  & -0.106775  & -0.106773  & -0.106771  \\
$E_{HF}$     & -2.456073  & -0.289341  &  0.003296  &  0.058799  &  0.059586  &  0.141323  &  0.200487  &  0.219569  &  0.220474  &  0.354558  &  1.309133  \\
\hline
number       &          1 &          2 &          3 &          4 &          5 &          6 &          7 &          8 &          9 &         10 &         11 \\
    1        & -1.0000000 &  0.0000000 &  0.0000001 &  0.0000003 & -0.0000005 & -0.0000003 & -0.0000002 &  0.0000018 & -0.0000003 &  0.0000003 &  0.0000000 \\
    2        &  0.0000001 &  0.9874431 & -0.1254053 & -0.0448357 &  0.0000086 & -0.0461695 &  0.0209189 &  0.0404288 &  0.0000061 &  0.0487713 &  0.0252404 \\
    3        &  0.0000001 &  0.1219484 &  0.9914155 & -0.0390963 &  0.0000074 & -0.0188644 &  0.0067139 &  0.0122555 &  0.0000018 &  0.0113715 &  0.0038543 \\
    4        & -0.0000001 &  0.0341548 &  0.0233464 &  0.7059131 &  0.7070066 & -0.0086162 &  0.0025685 &  0.0045386 &  0.0000007 &  0.0037416 &  0.0011241 \\
    5        &  0.0000006 &  0.0341581 &  0.0233485 &  0.7057122 & -0.7072069 & -0.0086170 &  0.0025687 &  0.0045391 &  0.0000007 &  0.0037419 &  0.0011242 \\
    6        & -0.0000003 &  0.0476655 &  0.0131934 &  0.0092906 & -0.0000018 &  0.9985814 &  0.0086809 &  0.0131154 &  0.0000020 &  0.0072567 &  0.0016799 \\
    7        &  0.0000003 &  0.0209372 &  0.0040808 &  0.0024014 & -0.0000005 &  0.0072696 & -0.9994546 &  0.0236000 &  0.0000034 &  0.0043975 &  0.0007768 \\
    8        & -0.0000010 &  0.0300438 &  0.0053228 &  0.0030247 & -0.0000006 &  0.0078266 & -0.0159271 & -0.7060887 & -0.7071987 &  0.0072404 &  0.0011349 \\
    9        & -0.0000015 &  0.0300426 &  0.0053226 &  0.0030245 & -0.0000006 &  0.0078263 & -0.0159267 & -0.7062729 &  0.7070148 &  0.0072401 &  0.0011348 \\
   10        & -0.0000003 &  0.0507207 &  0.0055301 &  0.0027759 & -0.0000005 &  0.0048671 & -0.0034514 & -0.0078920 & -0.0000012 & -0.9986423 &  0.0021888 \\
   11        & -0.0000000 &  0.0257544 &  0.0007582 &  0.0003350 & -0.0000001 &  0.0004543 & -0.0002459 & -0.0005020 & -0.0000001 & -0.0008708 & -0.9996673 \\
\hline
\end{tabular}
\end{sidewaystable}
\clearpage

\bibliographystyle{plainnat}
\nobibliography{RDMFT_KS,RDMFT_KS_alt}